\def\spose#1{\hbox to 0pt{#1\hss}}

\def\multleft#1{\hbox to size{\vbox {\halign {\lft{##}\cr #1}}\hfill}\par}
\def\multright#1{\hbox to size{\vbox {\halign {\rt{##}\cr #1}}\hfill}\par}

\def\today{\ifcase\month\or January\or February\or March\or April\or May\or
      June\or July\or August\or September\or October\or November\or December\fi
      \space\number\day, \number\year}

\def\Msun{\hbox{$\rm\thinspace M_{\odot}$}}

\def\H2{\hbox{H$_{2}$}}

\newcommand{\gtsim}{\mbox{{\raisebox{-0.4ex}{$\stackrel{>}{{\scriptstyle\sim}}
$}}}}
\newcommand{\ltsim}{\mbox{{\raisebox{-0.4ex}{$\stackrel{<}{{\scriptstyle\sim}}
$}}}}
\documentstyle[epsfig]{mn}
\include{defn}

\begin{document}
\hsize=6truein
       
\title[Radio galaxies spanning three decades in radio
luminosity: host galaxies and black-hole masses]
{A sample of radio galaxies spanning three decades in radio 
luminosity - I. The host-galaxy properties and black-hole masses}

\author[R.J.~McLure, et al.]
{Ross J. McLure$^{3,1}$\thanks{Email: rjm@roe.ac.uk}, Chris J. Willott$^{3,2}$,
Matt J. Jarvis$^{5,3}$, Steve Rawlings$^{3}$,\and Gary J. Hill$^{4}$, 
Ewan Mitchell$^{3}$, James S. Dunlop$^{1}$, Margrethe Wold$^{6}$\\
\footnotesize\\
$^{1}$Institute for Astronomy, University of Edinburgh, 
Royal Observatory, Edinburgh, EH9 3HJ, UK \\
$^{2}$National Research Council, 5071 West Saanich Rd, Victoria, B.C.,
V9E 2E7, Canada\\
$^{3}$Astrophysics, Department of Physics, Keble Road, Oxford, OX1
3RH, U.K.\\
$^{4}$McDonald Observatory, University of Texas at Austin, RLM 15.308,
Austin, TX 78712-1083, USA\\
$^{5}$Sterrewacht Leiden, Postbus 9513, 2300 RA Leiden, The
Netherlands\\
$^{6}$European Southern Observatory, Karl-Schwarzchild-Str. 2, D-85748
Garching, Germany}
\maketitle

\begin{abstract}
The results of analysis of HST $I-$band imaging of a sample of 
41 radio galaxies spanning three orders of magnitude in radio
luminosity at redshift $z\simeq0.5$ are presented. These results 
represent the
first stage in a coordinated programme to study the connection between
radio luminosity and host-galaxy properties, black-hole masses and 
cluster environments in radio galaxies spanning a wide range in radio
luminosity over a restricted range in redshift. The full sample is comprised of
objects drawn from four complete, low-frequency selected radio
samples with progressively fainter flux-density limits (3CRR, 6CE, 7CRS and 
the new TexOx-1000 sample). Modelling of the HST 
imaging data shows that the host galaxies have surface-brightness
distributions consistent with those expected for classic ellipticals 
(S\'{e}rsic parameter, $\beta\simeq0.25$), with $\beta$ in the
range $0.17<\beta<0.30$, and a mean of $<\beta>=0.23\pm0.01$. 
The luminosities of the host galaxies are found to be
comparable with those of galaxies drawn from the bright end
of the local cluster galaxy luminosity function, spanning the 
range $0.7L^{\star}<L<10 L^{\star}$, with a mean of $3.2\pm0.3
L^{\star}$, after correcting for the effects of passive evolution. 
In addition, the radio galaxies are shown to follow a 
Kormendy ($\mu_{e}-r_{e}$) relation indistinguishable from that of both
powerful low-redshift radio galaxies and local Abell brightest cluster
galaxies. Combining our new results with those in the literature it is found 
that the scalelengths and Kormendy relations of 3C-class radio
galaxies do not vary significantly over the redshift range $0.0<z<0.8$, 
providing no evidence for dynamical evolution of this class of host
galaxy within this redshift interval. Converting the host-galaxy 
luminosities into black-hole mass 
estimates, using the local $M_{bh}-M_{bulge}$ 
correlation, predicts that the radio galaxies harbour central 
black holes with masses in the range 
$10^{8.1}\Msun <M_{bh}< 10^{9.5}\Msun$, with a geometric mean of
$<M_{bh}>=10^{8.87\pm 0.04}\Msun$.  Finally, a significant ($\simeq 3\sigma$)
correlation is found between black-hole mass and 151-MHz radio
luminosity for those objects in the sample with either high-excitation
nuclear spectra (HEG) or classical double (CD) radio structures. 

\end{abstract}


\section{INTRODUCTION}
The underlying mechanisms which govern the large range in
low-frequency radio luminosity observed in radio-loud active galactic
nuclei (AGN) are currently unknown. Over the previous two decades a 
wealth of observational effort has
been invested in studying the redshift r\'{e}gime
$0<z<1$, investigating whether the dominant factors are linked
to the Mpc-scale cluster environments (eg. Prestage \& Peacock 1988;
Hill \& Lilly 1991; Ellingson, Yee \& Green 1991; Wold et al. 2000; 
Best 2000) or the kpc-scale properties of the 
host galaxies (e.g. Smith \& Heckman 1989; 
Best, Longair \& R\"{o}ttgering 1998; McLure \& Dunlop 2000; 
Dunlop et al. 2003; Zirm, Dickinson \& Dey 2003). 

One of the incontrovertible observational facts which has emerged
from these studies, at least at $z<1$, is that the host galaxies of all 
powerful radio-loud AGN are massive $L>L^{\star}$ ellipticals 
(eg. Taylor et al. 1996; McLure et al. 1999; Dunlop et
al. 2003). The apparent uniformity of radio-loud AGN host galaxies has
taken on added importance over the last few years, following the 
discovery in nearby (distance $\ltsim$ 150 Mpc) inactive galaxies that a
reasonably accurate estimate ($\Delta M_{bh}\simeq 0.3$ dex) of the 
central black-hole mass can be obtained via its correlation with the 
mass of the host spheroidal component 
(Kormendy \& Richstone 1995; Magorrian et al. 1998; Gebhardt et
al. 2000; Ferrarese \& Merritt 2000; McLure \& Dunlop 2002; 
Marconi \& Hunt 2003; Tremaine et al. 2002). Moreover, recent progress has 
also indicated that a similarly accurate black-hole mass estimate 
($\Delta M_{bh}\simeq 0.4$ dex) can be obtained for broad-line AGN using
emission-line widths to derive the virial mass estimate
(eg. Kaspi et al. 2000; McLure \& Dunlop 2002; McLure \& Jarvis 2002; 
Vestergaard 2002). Consequently, a
large body of work has appeared in the recent literature investigating
the possible link between radio luminosity and black-hole mass in
radio-loud AGN (eg.  Laor 2000; Lacy et al. 2001; McLure \& Dunlop
2001a; McLure \& Dunlop 2002; Bettoni et al. 2003; Dunlop et al. 2003).

Unfortunately, observational studies have traditionally been 
subject to a degeneracy between radio luminosity and redshift 
produced as a by-product of flux-limited radio samples. In order to 
study the properties of radio-loud AGN
separated by a large dynamic range in radio luminosity, it has
previously been necessary to select samples
consisting of objects covering a wide range of redshifts. This has led
to difficulties in interpreting the data due to the complication of
potentially significant evolutionary effects. However, by selecting
our sample of objects from four complete, low-frequency selected radio
samples with successively fainter flux-density limits, it has been
possible to construct a sample of radio galaxies which spans three 
decades in radio luminosity at a virtually constant cosmic 
epoch ($0.4<z<0.6$). In this respect the motivation for this study is
identical to that pursued by Hill \& Lilly (1991), who investigated the
cluster environments of a sample of 45 radio galaxies spanning a
similar range in radio luminosities at $z\simeq 0.5$. By successfully
disentangling the effects of luminosity and redshift for the first
time, Hill \& Lilly (1991) were able to demonstrate that $z\simeq 0.5$
marks an apparent epoch-dependent change in 
cluster environment, with radio galaxies
tending to inhabit significantly richer cluster environments than
their low-redshift counterparts. Furthermore, the wide range of radio
luminosities in the Hill \& Lilly sample enabled them to determine
that extended radio luminosity did not appear to be a strong function
of cluster richness.

Developments over the intervening decade have provided our new study with
two crucial advantages over the previous work of Hill \& Lilly
(1991). Firstly, the availability of the 7CRS (Willott et al. 2003;
Lacy et al. 1999) and 
TexOx-1000 (Hill \&  Rawlings 2003) radio samples has now allowed us to
compile a sample of 41 radio galaxies at $z\simeq 0.5$, which is
complete over the full three-decades range in radio power. Secondly, 
with the high-resolution imaging provided by the Hubble
Space Telescope (HST), and the improvement in follow-up observations
available with ground-based telescopes, we are now in a position to obtain
a wide range of high quality data-sets for our new sample which were
not obtainable only ten years ago. Consequently, our new sample of
$z\simeq 0.5$ radio galaxies is the basis for an ambitious project, which
will for the first time systematically 
investigate the connection between the radio
luminosity of radio galaxies, the properties of their 
host galaxies, their location on the 
fundamental plane, the richness and mass of their 
cluster environments and the masses of their central black holes.

The results presented in this paper represent the first stage of this
wide-ranging project, and concentrate solely on the analysis
of our deep $I-$band HST WFPC2 imaging data. The results of our 
extensive follow-up observations of the sample are deferred to a 
series of future papers.

The structure of the paper is as follows, in Section 2 the details of 
the radio-galaxy sample are described, while in Section 3 the HST
observations and data reduction are discussed. In Section 4 the
modelling of the HST imaging is described, with the main results
presented and compared to literature results on low-redshift radio
galaxies in Section 5. In Section 6 the host-galaxy properties of 
our 3CRR sub-sample are compared to literature results to explore 
the evidence for dynamical evolution of the most powerful radio
galaxies within the redshift range $0.0<z<0.8$. In Section 7 the 
properties of the radio-galaxy
hosts are compared to those of local brightest cluster galaxies. In
Section 8 the masses of the radio
galaxies' central black holes are estimated, and in Section 9 the
relationship between black-hole mass and radio luminosity is
investigated. The main conclusions are summarized in Section 10.
Unless otherwise stated, throughout this paper the following cosmology 
is assumed: $H_{0}=70$ km\,\,s$^{-1}$Mpc$^{-1}$, $\Omega_{m}=0.3$, 
$\Omega_{\Lambda}=0.7$.

\begin{table*}
\begin{center}
\caption{The full ZP5 sample. Column one lists the radio-galaxy 
names, while columns two and three list the J2000 source 
coordinates. Column 4 lists the radio-galaxy redshifts and 
column five lists the logarithm of the 151-MHz luminosities 
in units of$~$WHz$^{-1}$sr$^{-1}$. Column 6 classifies the 
radio structures as either Classical Double (CD), Fat Double (FD),
Jet (J) or Compact (CM), the same classifications as used by 
Owen \& Laing (1989). The radio structure classifications are based on
consistent 1.4 GHz VLA A-array data (1.5 arcsec resolution) for the 6CE,
7CRS and TOOT sub-samples. For the 3CRR sub-sample, for which
structural classifications are well established, the classifications 
were based on the high-quality maps available in the
literature (at a variety of frequencies/resolutions). Column 7 describes 
the optical nuclear spectra in terms of the presence
of high-excitation narrow emission lines (HEG) or not (LEG), using the
classification scheme of Jackson \& Rawlings (1997). Those objects
for which the available optical spectra are of insufficient quality to
allow a reliable HEG/LEG determination are listed with a `?'. To avoid
confusion we include here in parenthesis the old names of the 7CRS
objects which have been adopted in previous papers: 
7C0213+3418 (5C6.63), 7C0219+3419 (5C6.201), 7C0219+3423 (5C6.214), 
7C0220+2952 (5C6.233), 7C0223+3415 (5C6.279), 7C0810+2650 (5C7.7).}
\begin{tabular}{lcccccl}
\hline
Source       &   RA & DEC &z   &  $L_{151}$ &Radio Structure&Class\\
\hline
 3C 16        & 00 37 45.39 & +13 20 09.6 &0.405&  26.82 &CD&HEG\\
 3C 19        & 00 40 55.01 & +33 10 07.3 & 0.482&  26.96&CD&LEG\\
 3C 46        & 01 35 28.47 & +37 54 05.7 & 0.437&  26.84&CD&HEG\\
 3C 172       & 07 02 08.32 & +25 13 53.9 & 0.519&  27.17&CD&HEG\\
 3C 200       & 08 27 25.38 & +29 18 45.5 & 0.458&  26.92&CD&LEG\\
 3C 225B      & 09 42 15.41 & +13 45 51.0 & 0.582&  27.50&CD&HEG?\\
 3C 228       & 09 50 10.79 & +14 20 00.9 & 0.552&  27.37&CD&HEG\\
 4C 74.16     & 10 14 14.84 & +74 37 37.4 & 0.568&  27.16&CD&HEG\\
 3C 244.1     & 10 33 33.97 & +58 14 35.8 & 0.428&  27.10&CD&HEG\\
 3C 274.1     & 12 35 26.64 & +21 20 34.7 & 0.422&  27.02&CD&HEG\\
 3C 295       & 14 11 20.65 & +52 12 09.0 & 0.464&  27.68&CD&HEG\\
 3C 330       & 16 09 35.01 & +65 56 37.7 & 0.550&  27.43&CD&HEG\\
 3C 341       & 16 28 04.04 & +27 41 39.3 & 0.448&  26.88&CD&HEG\\
 3C 427.1     & 21 04 07.07 & +76 33 10.8 & 0.572&  27.53&CD&LEG\\
 3C 457       & 23 12 07.57 & +18 45 41.4 & 0.428&  27.00&CD&HEG\\
\hline
 6C 0825+3407 &  08 25 14.59 & +34 07 16.8 &0.406&  26.09&FD&LEG\\
 6C 0850+3747 &  08 50 24.77 & +37 47 09.1 &0.407&  26.15&FD&HEG\\
 6C 0857+3945 &  08 57 43.56 & +39 45 29.0 &0.528&  26.34&CD&HEG\\
 6C 1111+3940 &  11 11 19.39 & +39 40 14.5 &0.590&  26.33&CD&LEG\\
 6C 1132+3439 &  11 32 45.74 & +34 39 36.2 &0.512&  26.33&CD&HEG\\
 6C 1200+3416 &  12 00 53.34 & +34 16 47.3 &0.530&  26.17&CD&LEG\\
 6C 1303+3756 &  13 03 44.26 & +37 56 15.2 &0.470&  26.29&CD&HEG\\
\hline
 7C 0213+3418 &  02 13 28.39 & +34 18 30.6 &0.465&  25.66&CD&LEG\\
 7C 0219+3419 &  02 19 15.89 & +34 19 43.2 &0.595&  26.13&FD&HEG?\\
 7C 0219+3423 &  02 19 37.83 & +34 23 11.2 &0.595&  25.98&J&HEG?\\
 7C 0220+2952 &  02 20 34.26 & +29 52 19.5 &0.560&  26.07&FD&LEG?\\
 7C 0223+3415 &  02 23 47.24 & +34 15 11.9 &0.473&  25.55&CD&HEG\\
 7C 0810+2650 &  08 10 26.10 & +26 50 49.1 &0.435&  25.58&CM&HEG\\
 7C 1731+6638 &  17 31 43.84 & +66 38 56.7 &0.562&  25.62&CD&HEG\\
 7C 1806+6831 &  18 06 50.16 & +68 31 41.9 &0.580&  26.36&CD&HEG\\
\hline
 TOOT 0009+3523 &  00 09 46.90 & +35 23 45.1 &0.439&  24.79&FD&LEG\\
 TOOT 0013+3459 &  00 13 13.29 & +34 59 40.6 &0.577&  25.75&FD&LEG\\
 TOOT 0018+3510 &  00 18 53.93 & +35 10 12.1 &0.416&  25.16&J&LEG\\
 TOOT 1255+3556 &  12 55 55.83 & +35 56 35.8 &0.471&  25.01&J&LEG\\
 TOOT 1301+3658 &  13 01 25.03 & +36 58 09.4 &0.424&  24.76&J&LEG\\
 TOOT 1303+3334 &  13 03 10.29 & +33 34 07.0 &0.565&  25.66&J&HEG\\
 TOOT 1307+3639 &  13 07 27.07 & +36 39 16.4 &0.583&  25.30&J&LEG\\
 TOOT 1309+3359 &  13 09 53.95 & +33 59 28.2 &0.464&  24.91&CM&HEG\\
 TOOT 1626+4523 &  16 26 48.50 & +45 23 42.6 &0.458&  25.03&FD&LEG\\
 TOOT 1630+4534 &  16 30 32.80 & +45 34 26.0 &0.493&  25.17&J&LEG\\
 TOOT 1648+5040 &  16 48 26.19 & +50 40 58.0 &0.420&  25.12&CM&LEG\\  
\hline
\end{tabular}
\label{tab1}
\end{center}
\end{table*}

\begin{figure*}
\centerline{\epsfig{file=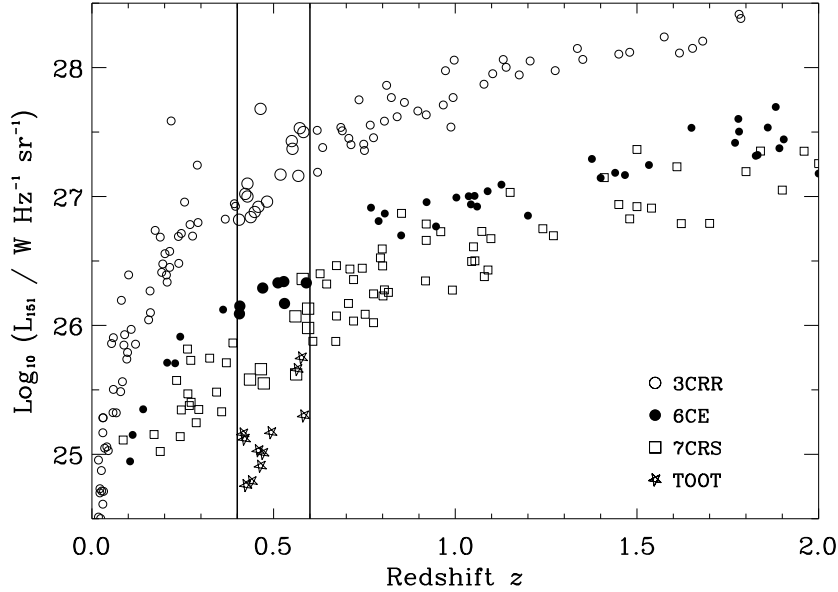,width=12cm,angle=0}}
\caption{151-MHz radio luminosity versus redshift for the 
narrow-line radio galaxy component of the 3CRR, 6CE and 7CRS
samples. The members of the ZP5 sample 
are highlighted as enlarged symbols in the
redshift interval $0.4<z<0.6$, including those drawn from the new 
TexOx-1000 (TOOT) survey.}
\label{fig1}
\end{figure*}

\begin{table*}
\begin{center}
\caption{List of sources which met the redshift and radio-luminosity
 criteria for inclusion in the ZP5 sample, but were excluded due to
 the presence of broad emission lines. In column 2 the sources are
 classified as either quasar (Q) or weak quasar (WQ) according to the
 classification scheme of Jackson \& Rawlings (1997). We note here
 that the excluded quasars are confined to the two most radio luminous
 sub-samples, in agreement with the findings of Willott et al. (2000)
 that the quasar fraction in complete, low-frequency selected radio 
 samples is an increasing function of radio luminosity. The
 implications of the possible quasar cores in the ZP5 sample (see
 Table 3) will be fully discussed in a forthcoming paper (Mitchell et
 al., in prep).}
\begin{tabular}{lcl}
\hline
Source         &Class&Comment\\
\hline
3C 47          &Q &\\
3C 147         &Q &\\
3C 215         &Q &\\
3C 275.1       &Q &\\
3C 334         &Q &\\
3C 455         &WQ&\\
6C 1220+3723   &Q & B1950.0 name - as in Rawlings et al. (2001)\\
5C 7.118       &WQ& B1950.0 name - as in Willott et al. (2003)\\
TOOT 1252+3310   &WQ& Broad H$\beta$ detected in follow-up spectroscopy
(Willott et al., in prep)\\
\hline
\end{tabular}
\label{tab2}
\end{center}
\end{table*}

\section{The sample}
\label{sample}

The full 41-object $z\simeq 0.5$ radio-galaxy sample (hereafter the ZP5 
sample) consists of all the narrow-line radio galaxies in the redshift
interval $0.4<z<0.6$ from four, complete, low-frequency selected radio
surveys; 3CRR (Laing, Riley \& Longair 1983), 6CE (Eales et al. 1997; 
Rawlings, Eales \& Lacy 2001), 7CRS (Willott et al. 2003; Lacy et
al. 1999) and 
TexOx-1000; hereafter TOOT (Hill \&  Rawlings 2003). The locations of 
all four radio surveys on the radio luminosity$~-~$redshift plane are 
shown in Fig \ref{fig1}. Details of the ZP5 sample can be found in
Table 1. Additionally, Table 2 lists those objects which satisfied the 
redshift and radio luminosity criteria for the ZP5 sample, but were 
excluded due to their broad-emission line spectra.

The TOOT objects were selected from a preliminary version of the 
survey. They are therefore a subset of all the TOOT sources meeting the
radio flux density and sky area selection criteria. Broad-line
objects were excluded, in the case of TOOT 1252+3310 after HST imaging 
and during further spectroscopic follow-up. There is some possibility that 
this subset is biased in the sense that the optically
faintest sources may be have been preferentially excluded, 
although the level of bias will only be clear once analysis of the 
full survey is complete.

The original decision to select the radio-galaxy sample in the
redshift interval $0.4<z<0.6$ was motivated by several
factors. Principal among these was that at $z\simeq 0.5$ it becomes possible 
to get a full three decades of dynamic range in radio luminosity,
including objects which are close to the most radio luminous objects
which exist. Secondly, at $z\simeq 0.5$ it is possible to construct
sub-samples, divided by radio luminosity, which contain sufficient
numbers of objects to reliably determine the mean values and variances
of relevant parameters, allowing any correlations with radio
luminosity to be investigated. Thirdly, the 
chosen redshift range makes it possible to obtain the deep, 
high-resolution images necessary to derive accurate
host-galaxy parameters, without requiring prohibitive amounts of HST
time, while the upper limit of $z=0.6$ ensures that with the WFPC2
filters we are still able to sample light longward of the 4000\AA\,break.
Finally, redshift $z\simeq0.6$ represents the
maximum redshift for which it is practicable on 4m-class telescopes 
to obtain accurate measurements of the host-galaxy stellar-velocity
dispersions, necessary to investigate the radio-galaxy fundamental 
plane properties (Willott~et al., in prep). 

\section{Observations and data reduction}
The observations for 39/41 of the objects in this project were 
carried out in the Cycle 10
program GO\#9045 with the Hubble Space Telescope. We used the WFPC2
detector with the F785LP filter. When combined with the system
efficiency this filter has a peak wavelength of 8700\,\AA\ and high
throughput from $7900$\,\AA $-$ $9300$\,\AA. This filter was selected because
we wish to be sampling light above the 4000\,\AA\, break, where there is
less likely to be contamination of the old-stellar population by more
recently formed stars or AGN-related emission. Further restrictions on
the filter choice came from our desire to be free from emission lines
as much as possible. The [O\,{\sc iii}]\,$\lambda5007$ line only
enters the filter bandpass at redshifts $z>0.58$ and H\,$\alpha$
leaves the filter bandpass at redshifts $z>0.42$. Since our sample
spans the redshift range $0.4<z<0.6$ this ensures very few sources
will have significant emission-line contamination with this filter.

The radio galaxies were placed in the WF3 chip so that information on
their arcminute-scale environments could be probed with the other
chips on WFPC2. The orientation of the detector on the sky was
arranged so that diffraction spikes and blooming from bright stars did
not affect the array at the locations of the radio galaxies. The
observations of each radio galaxy were made during a single orbit and
the total exposure time of 2000 seconds was split into 4 integrations
of 500s each to facilitate the removal of cosmic rays. The target was
dithered 5.5 WF3 pixels in $x$ and $y$ in between integrations to
improve the sampling of the PSF.

Calibrated images were retrieved from the HST archive. Registering and
combining the 4 images per target were performed using the {\sc drizzle}
routines in the {\sc iraf} package (Fruchter \& Hook 2002). The final 
reduced images for all four CCDs were sampled on a grid of pixels with size 0.0498
arcsec. Unsaturated stars typically have FWHM values of 3 pixels
corresponding to 0.15 arcsec. 

Existing observations were used in the analysis of the two remaining 
objects in the sample; 3C16 and 3C295. In the case of 3C16, an
existing HST WFPC2 observation using the PC chip and the F702W filter
was obtained from the HST archive (GO\#6675). The pixel size of the PC
chip (0.045 arcsec/pixel) is well matched to that of the final 
drizzled images of the rest of the sample, and because 3C16 is at the
low-redshift end of the sample ($z=0.405$), the F702W filter is still
sufficiently red to sample light longward of the 4000\AA\
break. Although an archival HST observation of 3C295 was also available,
because of the placement of the radio galaxy close to the edge of the
PC chip, it was deemed unsuitable for the host-galaxy analysis
described in Section~\ref{anal}. Fortunately we were provided 
with a deep $I-$band image of the
3C295 cluster obtained on the 2.5m Nordic Optical Telescope for the 
weak lensing analysis of Wold et al. (2002). The final image comprises
10,000 seconds of integration and has a FWHM of 0.8 arcsecs. Although
this is clearly lower spatial resolution than the HST data, this does
not have any influence on the accurate determination of the host-galaxy
properties because 3C295 is the largest, most luminous target in the
whole sample. 

\section{Host-galaxy analysis}
\label{anal}
\begin{figure*}
\centerline{\epsfig{file=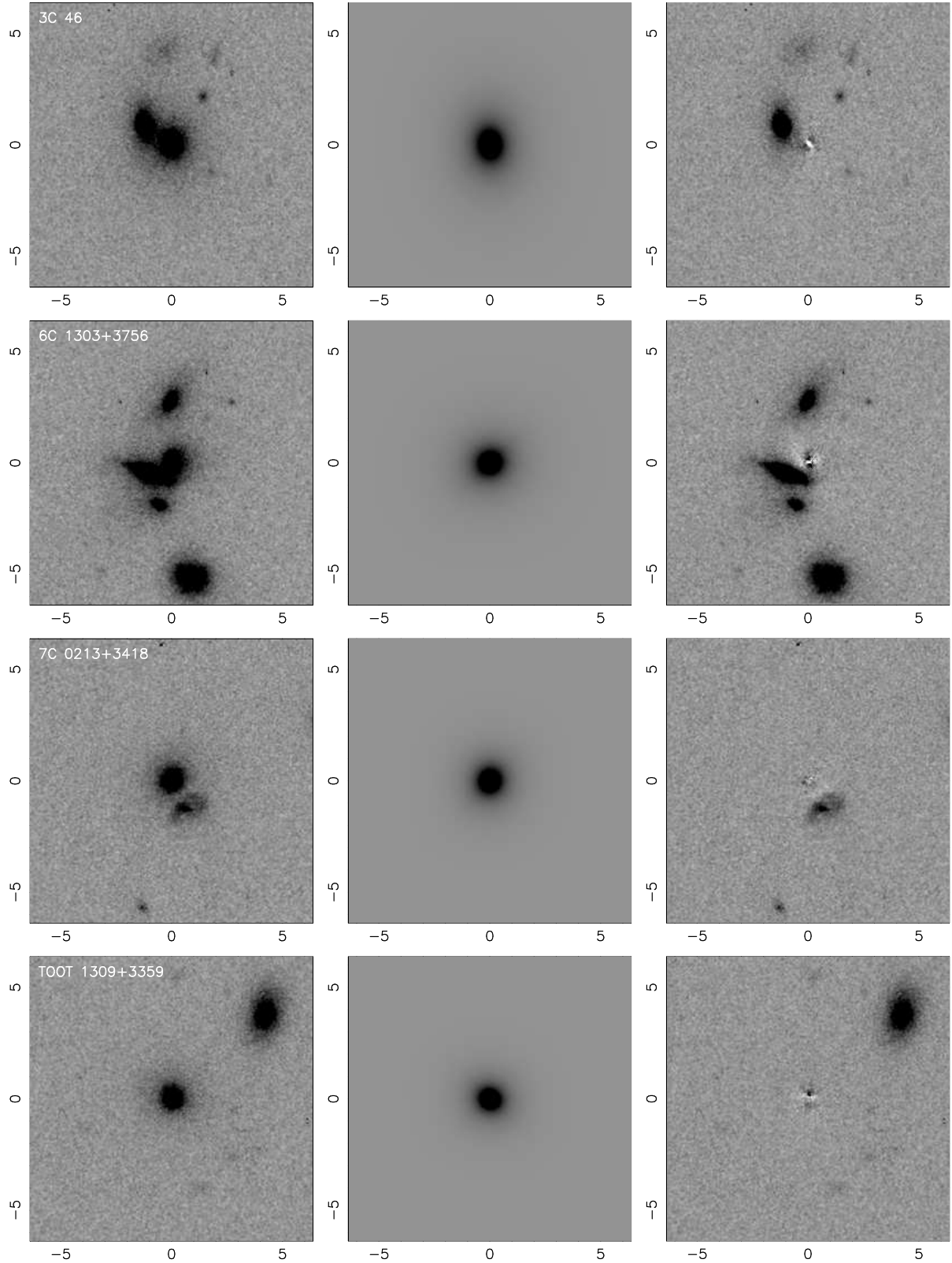,width=16cm,angle=0}}
\caption{Examples of the two-dimensional modelling analysis of the
HST data for one object from each of the four sub-samples. The
left-hand panels show cut-outs of the reduced HST WFPC2 images centred 
on the radio galaxies. The middle panels show the best-fitting 
host-galaxy models. The right-hand panels show the residual images 
produced by subtracting the best-fitting models from the reduced
radio-galaxy images. All panels are $12.75''\times12.75''$.}
\end{figure*}

Given the high resolution and depth of the HST data available
for this study it was decided that the determination of the
host-galaxy parameters should be carried out via full
two-dimensional modelling of the images. The
two-dimensional modelling technique adopted is identical to that which
has previously been applied in the analysis of both radio galaxies 
and quasar host galaxies (eg. McLure et al. 1999; McLure \&
Dunlop 2000; Dunlop et al. 2003). Full details of this modelling
technique have been previously published in McLure, Dunlop \& Kukula
(2000) and so only a brief outline will be provided here. 

The first stage in the process is the production of a two-dimensional 
model host galaxy which has a surface-brightness distribution governed
by an axially-symmetric S\'{e}rsic (1968) profile of the form:
\begin{equation}
\Sigma(r)=\Sigma_{0} \exp\left[ -\left(\frac{r}{r_{0}}\right)^{\beta} \right]
\end{equation}
\noindent
where $\Sigma_{0}$ is the central surface brightness, $r_{0}$ is the
characteristic scalelength and the $\beta$ parameter controls the
shape of the radial profile. Under the adopted form of the  S\'{e}rsic
profile a value of $\beta=0.25$ corresponds to a classical de Vaucouleurs
$r^{1/4}$ law (de Vaucouleurs 1953) and a value of $\beta=1.0$ 
corresponds to an exponential  Freeman (1970) disc model. This model galaxy is
subsequently convolved with the HST point spread function in order to
produce a synthetic host-galaxy image. A chi-square comparison of the data
and model is then performed on a pixel-by-pixel basis, in order to
determine the goodness of fit of the initial model. A chi-square
minimization process is then performed to determine the best-fitting
value of the six parameters which describe the model ($\Sigma_{0},
r_{0}, \beta$, axial ratio, position angle and nuclear point-source 
luminosity). Although the model galaxy axial ratio and position angle
are included as free parameters during the fitting process, they are
held constant as a function of radius. During the model fitting 
process it was not assumed {\it a priori} that the radio 
galaxies would not possess an unresolved nuclear point source.

\subsection{Point spread function}
For the analysis of the HST imaging data we utilized the model WFPC2
point spread functions (PSF) produced by the TinyTim software package
(TinyTim v6.0, Krist 2001). The great advantage of using the TinyTim
model PSFs over empirical WFPC2 PSFs is that they have infinite
signal-to-noise, and can be produced to match the $(x,y)$ coordinates of each radio
galaxy on the WF3 chip. Although the model PSFs produced by TinyTim do not
reproduce the large-scale halo of scattered light of the empirical
WFPC2 PSF at radii of $r>1.5''$, this was not considered to be a
serious problem for the current analysis because the luminosity of the
any nuclear point sources were anticipated to be only a small
fraction of the host-galaxy luminosity. This assumption proved to be
justified, with the vast majority of the radio
galaxies proving to possess a negligible nuclear point-source
component (the median nuclear:host flux ratio for the ZP5 sample is $0.03$).

In order to match the final reduced radio-galaxy images as accurately as
possible, model PSFs were produced at the two dithered positions of
each radio-galaxy observation separately. The two model PSFs were then 
processed by the {\sc drizzle} routine in an identical fashion to the
radio-galaxy images to produce the final model PSF used during the 
modelling process.

\section{Results}
\label{results}
In this section the results of the two-dimensional modelling of the
host galaxies of the ZP5 sample are presented and discussed. A list of
the principal host-galaxy parameters derived from the model fits for
each individual object can be found in Table \ref{tab3}, while the
average values for parameters of particular interest for the
four radio-galaxy sub-samples and the full ZP5 sample can be found in
Table \ref{tab4}. A graphical representation of the modelling process
is shown in Fig 2 which, for one radio galaxy from
each sub-sample, shows postage-stamp images of the reduced HST data,
the best-fitting two-dimensional models and the residual images after 
subtraction of the best-fitting models.
\begin{center}
\begin{table*}
\caption{Host-galaxy parameters derived from the two-dimensional 
modelling of the HST WFPC2 data. Column one lists the 
radio-galaxy names. Column 2 lists the effective radii in kpc. Column 3
lists the mean $R-$band surface brightness within the effective radii in
mag.arcsec$^{-2}$. The surface brightnesses have been corrected for
$(1+z)^4$ cosmological dimming, and then K-corrected and
corrected for passive evolution as described in the text. Column four 
lists the axial ratios and column five
lists the real optical position angles of the radio galaxies in
degrees, transformed from the fitted on-chip position angles. Column six 
lists the S\'{e}rsic $\beta$ parameters. Column 7 lists the integrated
absolute $R-$band host luminosities, including K-correction and
correction for passive evolution as described in the text. Column 8
lists the host-galaxy luminosities in units of $L^{\star}$, 
where $L^{\star}$ is taken to 
be $M_{R}=-22.07$ (Yagi et al. 2002). Column 9 lists the ratio of
apparent nuclear:host luminosity. Column 10 lists the estimated
black-hole mass for each object based on its fitted bulge
luminosity as described in Section 8. Column 11 highlights properties
of the radio galaxies revealed by the model-subtracted images:
UD=undisturbed, M=merger, MM=major merger, DA?=possible disc/aligned
emission, QC?=possible quasar core (see text for discussion).}
\begin{tabular}{lccclcccccl}
\hline
Source&$r_{e}$&
$<\mu>_{e}$&$a/b$&PA&$\beta$&$M_{R}$&$L/L^{\star}$&$L_{nuc}/L_{host}$&$\log(M_{bh}/\Msun$)&Comments\\
\hline
3C 16       &  22.9&  21.75& 1.43 & 152& 0.20& $-$23.23&2.9&0.028&8.87&DA?\\
3C 19       &  30.9&  21.86& 1.31 &\phantom{0}55& 0.21&$-$23.81&5.0&0.019&9.17 &M/DA?\\
3C 46       &  15.8&  20.78& 1.32&   129& 0.21& $-$23.49       &  3.7&0.002&9.00&MM\\
3C 172      &  12.6&  20.65& 1.18& \phantom{00}4& 0.23& $-$23.25&  3.0&0.011&8.89&UD\\
3C 200      &  13.2&  20.15& 1.68& 141& 0.26& $-$23.48          &  3.6&0.059&9.00&DA?\\
3C 225B     &  18.6&  21.04& 1.42& 115& 0.23& $-$23.49          &  3.7&0.025&9.00&UD\\
3C 228      &  13.2&  21.17& 1.38& 105& 0.25& $-$22.64          &  1.7&0.113&8.58&M/QC?\\
4C 74.16    &  18.6&  21.64& 1.28&\phantom{0}48& 0.20&$-$23.04&    2.5&0.065&8.78&DA?\\
3C 244.1    &  15.8&  20.76& 1.40&\phantom{0}82& 0.25& $-$23.45 &  3.5&0.060&8.98&DA?\\
3C 274.1    &  \phantom{0}9.5&  20.20& 1.09& 116& 0.26& $-$23.18&  2.8&0.000&8.85&UD\\
3C 295      &  29.5&  21.03& 1.44& 136& 0.29& $-$24.52       &9.5&0.000&9.52     &MM \\
3C 330      &  20.9&  21.26& 1.21& \phantom{0}21& 0.20& $-$23.71&  4.6&0.081&9.11&MM\\
3C 341      & 16.6&  21.08& 1.16& 115& 0.20& $-$23.45        &  3.5&0.069&8.99   &DA?\\
3C 427.1    &  18.2&  20.68& 1.42& \phantom{0}31& 0.25& $-$23.81&  5.0&0.000&9.16&UD\\
3C 457      &  14.1&  20.40& 1.28& \phantom{0}14& 0.22& $-$23.67&  4.4&0.013&9.09&DA?\\
\hline
6C 0825+3407&  14.5&  21.12& 1.09& 112& 0.27& $-$23.14        &  2.7&0.000&8.83&M\\
6C 0850+3747&  12.6&  20.41& 1.34& 160& 0.24& $-$23.35          &  3.2&0.131&8.93&UD/QC?\\
6C 0857+3945&  11.7&  20.15& 1.34& \phantom{00}6& 0.22&  $-$23.42& 3.5&0.011&8.97&UD\\
6C 1111+3940&  \phantom{0}5.2&  19.63& 1.06& 174& 0.25& $-$22.48          &  1.4&0.290&8.65&UD/QC?\\
6C 1132+3439&  10.5&  20.40& 1.35& \phantom{0}10& 0.22& $-$22.96& 2.3&0.028&8.74&UD\\
6C 1200+3416&  20.0&  21.55& 1.15&\phantom{0}50& 0.17& $-$23.35& 3.2&0.000&8.93&MM\\
6C 1303+3756&  12.3&  20.65& 1.09& 106& 0.20& $-$23.28         &  3.1&0.067&8.90&MM/DA?\\
\hline
7C 0213+3418&  \phantom{0}6.3&  19.47& 1.09& \phantom{0}69& 0.24& $-$22.99&  2.3&0.000&8.75 &UD\\
7C 0219+3419&  10.9&  20.45& 1.53& 124& 0.26& $-$22.87          &  2.1&0.042&8.69           &UD\\
7C 0219+3423&  \phantom{0}7.8&  20.67& 1.08&  132& 0.25& $-$22.26         &  1.2&0.031&8.39 &UD\\
7C 0220+2952&  12.9&  21.58& 1.09&\phantom{0}87& 0.24& $-$22.43 &  1.4&0.023&8.47           &UD\\
7C 0223+3415&  \phantom{0}9.8&  20.51& 1.42&  139& 0.22& $-$22.62         &  1.7&0.047&8.57 &UD\\
7C 0810+2650&  14.8&  21.16& 1.05&\phantom{0}62& 0.23& $-$23.22 &  2.9&0.080&8.87           &UD\\
7C 1731+6638&  \phantom{0}6.0&  20.47& 1.36&\phantom{0}48& 0.22& $-$21.67 &0.7&0.268&8.10   &UD/QC?\\
7C 1806+6831&  14.8&  20.54& 1.21& 148& 0.19& $-$23.68        &  4.4&0.000&9.10             &M\\

\hline
TOOT 0009+3523&  12.6&  20.48& 1.97&\phantom{0}86& 0.29& $-$22.87&  2.1&0.037&8.70  &MM\\
TOOT 0013+3459&  15.5&  19.85& 1.64&  128& 0.30& $-$24.14       &  6.8&0.000&9.33   &M\\
TOOT 0018+3510&  19.5&  21.21& 1.32& \phantom{0}49& 0.20& $-$23.51&3.8&0.039&9.01 &DA?\\
TOOT 1255+3556&  \phantom{0}7.8&  20.42& 1.07& \phantom{00}2& 0.25& $-$22.53&  1.5&0.155&8.52 &UD/QC?\\
TOOT 1301+3658&  21.9&  21.17& 1.29& 140& 0.20& $-$23.79    &  4.9&0.076&9.16 &DA?\\
TOOT 1303+3334&  \phantom{0}5.1&  19.69& 1.26&\phantom{0}16& 0.21& $-$22.15&  1.1&0.743&8.42 &DA?/QC?\\
TOOT 1307+3639&  12.0&  20.31& 1.12& 110& 0.22& $-$23.51           &  3.8&0.084&9.01 &DA?\\
TOOT 1309+3359&  \phantom{0}6.6&  19.84& 1.11& \phantom{0}56& 0.21& $-$22.70&  1.8&0.025&8.61 &DA?\\
TOOT 1626+4523&  \phantom{0}7.4&  19.54& 1.19& \phantom{0}70& 0.23& $-$23.19 &  2.8&0.060&8.86 &DA?\\
TOOT 1630+4534&  10.2&  19.71& 1.03&  126& 0.23& $-$23.88        &  5.2&0.028&9.20 &MM\\
TOOT 1648+5040&  \phantom{0}9.5&  19.71& 2.09&\phantom{00}1& 0.27& $-$22.97&  2.3&0.000&8.75 &DA?\\
\hline
\end{tabular}
\label{tab3}
\end{table*}
\end{center}
\begin{center}
\begin{table*}
\caption{Average values of the host-galaxy parameters derived from the
two-dimensional modelling for the four radio-galaxy
sub-samples and the ZP5 sample as a whole. For each parameter the mean
value for the relevant sample is listed with its associated standard
error, and the median value is listed immediately below. Column 2
lists the S\'{e}rsic $\beta$ parameters. Column 3 lists the axial
ratios, while column 4 lists the effective scalelengths in kpc. Column
5 lists the mean $R-$band surface brightness within the effective radius in
mag.arcsec$^{-2}$. Column 6 lists the absolute integrated $R-$band
magnitudes, while column 7 lists the host-galaxy luminosities in units
of $L^{\star}$.}
\begin{tabular}{lcccccc}
\hline
Sample& $\beta$ & $a/b$& $r_{e}$ & $<\mu>_{e}$ & $M_{R}$ & $L/L_{\star}$ \\
\hline
3CRR & $0.23$$\pm0.01$          &$1.33$$\pm0.04$           &$18.1$$\pm1.5$          &$20.96$$\pm0.14$          &$-23.48$$\pm0.11$             &$3.95$$\pm0.46$\\
     & $0.23$$\phantom{\pm0.03}$&$1.32$$\phantom{\pm0.15}$ &$16.8$$\phantom{\pm2.7}$&$21.03$$\phantom{\pm0.67}$&$-23.48$$\phantom{\pm0.48}$&$3.65$$\phantom{\pm2.22}$\\
\hline
6CE &  $0.23$$\pm0.01$ & $1.20$$\pm0.05$ & $12.4$$\pm1.6$ & $20.56$$\pm0.24$& $-23.14$$\pm0.12$& $2.78$$\pm0.27$\\
    &  $0.23$$\phantom{\pm0.04}$&$1.15$$\phantom{\pm0.13}$&$12.3$$\phantom{\pm1.8}$&$20.41$$\phantom{\pm0.64}$&$-23.28$$\phantom{\pm0.25}$&$3.06$$\phantom{\pm0.60}$\\
\hline
7CRS & $0.23$$\pm0.01$ & $1.23$$\pm0.07$ & $10.4$$\pm1.3$ & $20.61$$\pm0.22$& $-22.72$$\pm0.22$& $2.08$$\pm0.41$\\
     & $0.24$$\phantom{\pm0.02}$&$1.15$$\phantom{\pm0.17}$&$9.4$$\phantom{\pm1.40}$&$20.53$$\phantom{\pm0.80}$&$-22.74$$\phantom{\pm0.64}$&$1.87$$\phantom{\pm1.13}$\\
\hline
TOOT & $0.25$$\pm0.01$ & $1.37$$\pm0.11$ & $11.7$$\pm1.6$ & $20.18$$\pm0.18$& $-23.20$$\pm0.19$& $3.28$$\pm0.54$\\
      &$0.23$$\phantom{\pm0.05}$&$1.26$$\phantom{\pm0.34}$&$10.3$$\phantom{\pm1.5}$&$19.85$$\phantom{\pm0.69}$&$-23.19$$\phantom{\pm0.55}$&$2.82$$\phantom{\pm1.60}$\\
\hline
FULL & $0.23$$\pm0.01$ & $1.30$$\pm0.04$ & $13.9$$\pm0.9$ & $20.61$$\pm0.10$& $-23.20$$\pm0.09$& $3.21$$\pm0.26$\\
   &   $0.23$$\phantom{\pm0.05}$&$1.28$$\phantom{\pm0.34}$&$12.8$$\phantom{\pm1.5}$&$20.54$$\phantom{\pm0.69}$&$-23.25$$\phantom{\pm0.55}$
&$2.97$$\phantom{\pm1.60}$\\
\hline
\end{tabular}
\label{tab4}
\end{table*}
\end{center}
\subsection{Morphology}

The mean value of the S\'{e}rsic $\beta$ parameter for the full ZP5
sample is $<\beta> =0.23\pm{0.01}$, consistent with that
expected for classical elliptical galaxies (the typical uncertainty in
$\beta$ is $\pm 0.05$). The average $\beta$ values
displayed in Table \ref{tab4} for the four radio-galaxy sub-samples
show no evidence for any trend between the $\beta$ parameter and 
radio luminosity. 

\begin{figure}
\begin{center}
\centerline{\epsfig{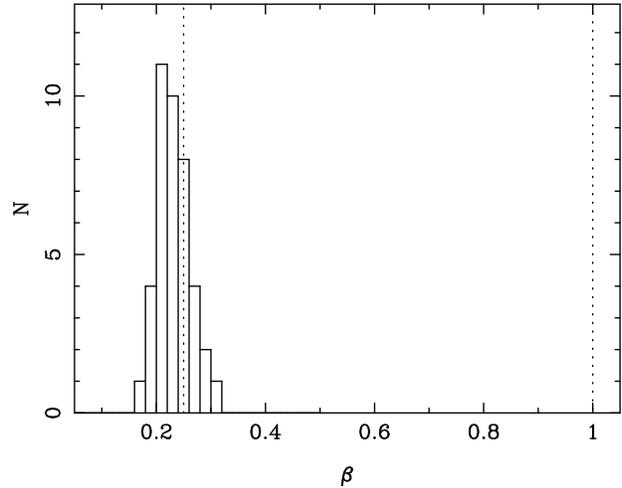}}
\caption{Histogram of the derived S\'{e}rsic $\beta$ parameters for the
full ZP5 sample. The dotted vertical lines indicate the locations of
$\beta=0.25$ and $\beta=1.0$, the expected values for classic de Vaucouleurs
$r^{1/4}$ and Freeman exponential disk profiles respectively (de
Vaucouleurs 1953; Freeman 1970).}
\label{fig3}
\end{center}
\end{figure}

In Fig \ref{fig3} we show a histogram of the derived $\beta$
parameters for the full ZP5 sample. Although Fig \ref{fig3} confirms
that the morphologies of the ZP5 sample radio galaxies are consistent 
with the classical $\beta=0.25$ value, we note here that there
are a substantial number of objects with $\beta<0.25$. The possible
implications of this in light of the recent discovery of a correlation
between $\beta$ and central black-hole mass (Graham et al. 2001; Erwin
et al. 2003) will be discussed in Section \ref{graham}.

The distribution of $\beta$ parameters derived for the
ZP5 sample is in good agreement with that determined by Dunlop et
al. (2003) for their sample of twenty 3C-class radio galaxies and radio-loud
quasars at $z\simeq0.2$. The mean $\beta$ parameter determined 
by Dunlop et al. for their radio-loud AGN sample was $<\beta>=0.24\pm0.01$. 
Unfortunately there are no 
comparable samples in the literature of radio galaxies at 
higher redshift, for which S\'{e}rsic $\beta$ parameters have been
derived, with which to compare the morphological results derived for
the present sample. 

\subsubsection{Model-subtracted images}

In the final column of Table 3 brief comments are listed describing
the structure revealed by subtracting the best-fitting
axially-symmetric host-galaxy model from the 
radio-galaxy images (see Fig 2
for examples). Objects for
which the best-fitting galaxy model produces a clean subtraction, with
little or no residual structure present in the
subtracted images, are described as being undisturbed (UD). These objects
comprise one third of the full ZP5 sample. Objects which show
clear signs of interaction are described as being mergers (M), with
those cases where the secondary nucleus/companion object has
a luminosity $>$~10\% of that of the radio-galaxy being described as major
mergers (MM). Objects for which the modelling indicates the presence
of an unresolved
point-source nucleus with luminosity $>$~10\% of that of the radio
galaxy (eg. TOOT~1303+3334, 7C~1731+6638) are described as possible
quasar cores (QC?). Finally, those objects
whose model-subtracted images reveal symmetric, central positive flux
residuals are described as having possible disc or aligned emission (DA?).

The information provided by the model-subtracted images is not pursued
further within this paper, which concentrates on the properties of the
dominant smooth stellar component of the radio-galaxy hosts. 
However, this information will be fully exploited by a forthcoming paper which
explores the link between radio luminosity/structure and the kpc-scale
environments of the radio galaxies, the incidence and
magnitude of interaction/merger activity, aligned emission components
and the properties of those radio galaxies with possible quasar cores 
(Mitchell et al., in prep). 

\subsection{Host-galaxy luminosity}
\label{lum}
\begin{figure}
\centerline{\epsfig{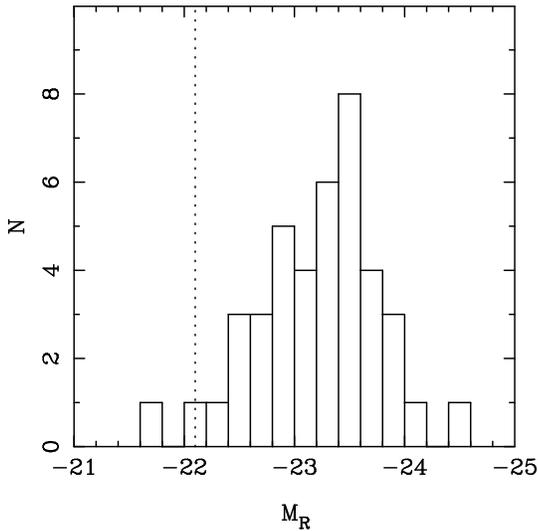}}
\caption{Histogram of integrated absolute $R-$band host magnitudes for
the full ZP5 sample. The vertical dotted line indicates the location
of $L^{\star}$ for local cluster galaxies: $M_{R}=-22.07$ (Yagi et
al. 2002).}
\end{figure}

The absolute magnitudes of the radio galaxies have
been calculated under the assumption that the spectral
energy distributions of the hosts can be well represented by the
passively evolving stellar populations typical of massive early-type
galaxies. This decision is based on several lines of evidence,
including the morphological results presented in the
previous section, the $R-K$ colour versus redshift distribution of 
7CRS objects (Willott, Rawlings \& Blundell 2001) and the 
off-nuclear spectroscopy of $z\simeq 0.2$ quasar host 
galaxies (Nolan et al. 2000). Furthermore,
the initial results from our follow-up spectroscopic
observations, designed to determine the host-galaxy stellar-velocity
dispersions, indicate that the host-galaxy spectra are consistent 
with mature stellar populations (Willott et al., in prep).

The absolute magnitudes for each host galaxy presented in Table
\ref{tab3} are based on the integrated instrumental (F785LP) 
magnitudes returned by the modelling code. These instrumental magnitudes
are then converted to apparent $I-$Cousins magnitudes using 
the filter transformations and colour equations detailed in 
Holtzman et al. (1995). The apparent colours required for this
transformation were calculated from the 1995 version of Bruzual \&
Charlot single-burst elliptical galaxy stellar-population models 
(Bruzual \& Charlot 1993), under the assumption of a $z=5.0$ formation 
redshift. Following correction for cosmological dimming within our 
chosen cosmology, evolution and
$K-$corrections were then applied using the same Bruzual \& Charlot 
models. Galactic extinction corrections 
were then applied to each object (Schlegel et al. 1998), before a final
conversion to absolute $R-$Cousins magnitudes was made, assuming a
$z=0$ colour of $R-I=0.7$ (Bruzual \& Charlot 1993), in order to 
facilitate a straightforward comparison with literature 
results on radio-galaxy properties. The surface-brightness results
presented in Tables 3 \& 4 have been corrected for $(1+z)^4$
cosmological dimming, and then K-corrected and corrected for passive
evolution in the same fashion as the absolute magnitudes.
\begin{figure}
\centerline{\epsfig{file=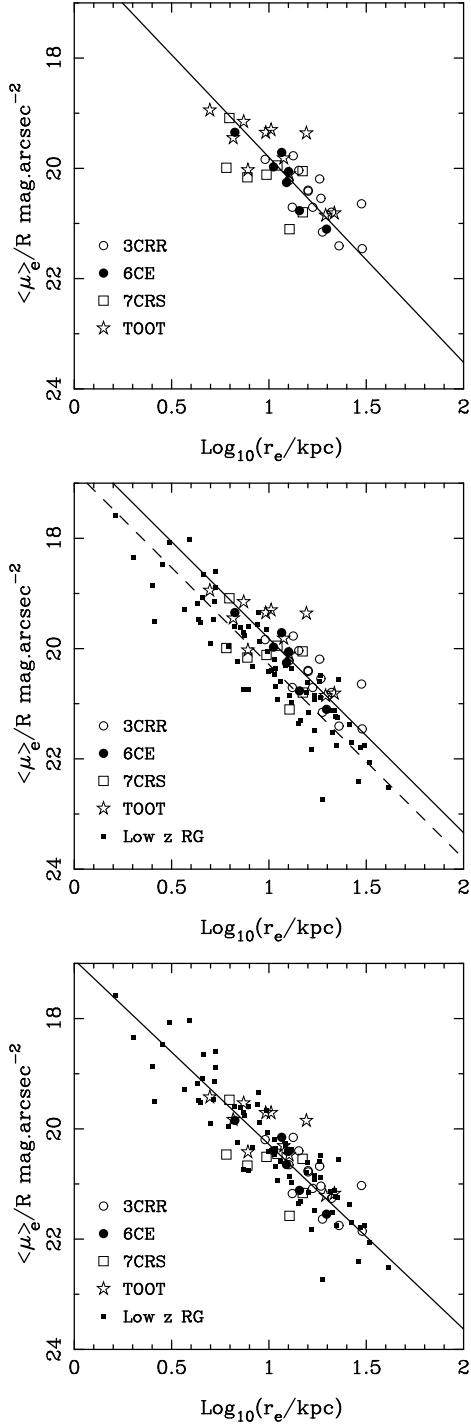,width=6.25cm,angle=0}}
\caption{The top panel shows the best-fitting Kormendy relation 
for the full ZP5 sample, before the application of any 
evolutionary corrections. The middle panel
shows the same ZP5 data with the addition of the 73 low-redshift radio
galaxies from Bettoni et al. (2001). The best-fitting fixed-slope Kormendy
relations to the ZP5 and Bettoni et al. radio galaxies are shown 
as the solid and dashed lines respectively (see text for discussion). 
The bottom panel shows the best-fitting Kormendy relation 
to the combined sample of the ZP5 and Bettoni et al. radio
galaxies. In the bottom panel the ZP5 host galaxies have been
corrected for passive evolution of their stellar populations as
described in the text.}
\label{fig5}
\end{figure}
The host-galaxy luminosities presented in Tables \ref{tab3}
\& \ref{tab4} and Fig 4 demonstrate that the host galaxies of all four
sub-samples are luminous, with a mean integrated $R-$band absolute
magnitude for the full ZP5 sample of $<M_{R}>=-23.20 \pm0.09$. In fact, these 
luminosities are directly comparable to the luminosities of galaxies 
drawn from the bright end of the local cluster-galaxy luminosity 
function. A comparison with the recent study of Yagi et
al. (2002) shows that the mean host-galaxy luminosity of the 
full ZP5 sample is $3.2\pm 0.3 L^{\star}$, where the Yagi et
al. determination of $L^{\star}$ for local cluster galaxies is
$M_{R}=-22.07$ in our adopted cosmology. 

The mean host-galaxy luminosities presented in Table \ref{tab4} also
allow a comparison to be made with the host-galaxy luminosity 
differences between the 3C, 6C and 7C samples determined from the
$K-z$ diagram (Willott et al. 2003; Eales et al. 1997). In the
$K-$band, Willott et al. find a mean luminosity difference of 0.33 
magnitudes between the 3C and 6C samples, and a corresponding
difference of 0.55 magnitudes between the 3C and 7C samples. The
equivalent figures from the analysis of our HST imaging are 0.34$\pm
0.16$ magnitudes and $0.76\pm 0.25$ magnitudes in the $R-$band, 
in good agreement with the $K-$band results.

The rough scaling between extended radio luminosity and host-galaxy
luminosity apparent from both the $K~-~z$ relation and the new results
presented here, suggests a correlation exists between radio luminosity and
black-hole mass within the 3CRR, 6CE and 7CRS samples. This issue
is discussed further in Section \ref{correlations}.

\subsection{Host-galaxy scalelengths}
\label{scale}

The best-fitting scalelength from the two-dimensional modelling of
each individual member of the ZP5 sample is listed in Table
\ref{tab3}, while the mean and median values for the four
radio-galaxy sub-samples and the full ZP5 sample are listed in Table
\ref{tab4}.

The full range of scalelengths displayed by the objects
in the ZP5 sample is $5<r_{e}<31$ kpc, with a mean value of $13.9
\pm 0.9$ kpc. Scalelengths of this size clearly place the radio
galaxies in the r\'{e}gime of brightest cluster
galaxies (eg. Graham et al. 1996; Hoessel, Oergerle \& Schneider
1987) and a  comparison of the ZP5 host galaxies with the properties of
brightest cluster galaxies (BCGs) is pursued further in 
Section \ref{bcg_comp}. Furthermore, in Section 6 the scalelength
results for the 3CRR sub-sample are combined with existing literature
results on 3C-class galaxies to search for any evidence for dynamical
evolution amongst the most powerful radio galaxies within the redshift
interval $0.0<z<0.8$.

In common with the results regarding the ZP5 host-galaxy luminosities,
an inspection of Table \ref{tab4} reveals an apparent correlation 
between mean scalelength and radio luminosity within the 3CRR, 6CE and 
7CRS sub-samples. This is entirely as expected
given the correlation between host-galaxy and radio luminosity
discussed above, and the well known correlation
between host luminosity and scalelength among early-type galaxies 
(e.g. $r_{e} \propto L_{host}^{0.63\pm0.03}$; Bernardi et al. 2003)

\subsection{The Kormendy relation}
\label{kormendy}

Early-type galaxies are known to exist on a two-dimensional manifold
(fundamental plane) in the three-dimensional parameter space 
defined by effective
scalelength, the mean surface brightness within the effective
scalelength and the central stellar-velocity 
dispersion (eg. Dressler, Lynden-Bell \& Burstein
1987; Djorgovski \& Davis 1987). At present we are engaged in 
obtaining stellar-velocity dispersion measurements for the ZP5 sample 
to investigate their location on the fundamental plane, and to 
estimate the masses of their central black holes 
via the $M_{bh}-\sigma$ correlation (Willott et al., in prep). 
However, armed with just the scalelength and surface
brightness parameters for each host galaxy, it is possible to examine the
photometric projection of the full fundamental plane, the so-called 
Kormendy or $\mu_{e}-r_{e}$ relation (Kormendy 1977).

The Kormendy relation for the ZP5 sample is shown in the top panel of Fig
\ref{fig5}. In this panel the $<\mu>_{e}$ values of the
ZP5 objects have been K-corrected and corrected for $(1+z)^4$
surface-brightness dimming, but have not been corrected for passive
evolution. Using the iterative $\chi^{2}$ FITEXY routine, which takes
account of errors in both parameters (Press et al. 1992), the
best-fitting form of the Kormendy relation for the non-evolutionary 
corrected ZP5 sample is:
\begin{equation}
<\mu>_{e}=3.72(\pm 0.51)\log r_{e} + 16.08(\pm 0.59)
\end{equation}
In the middle panel of 
Fig \ref{fig5} the same ZP5 data are repeated, along
with the 73 low-redshift radio galaxies from the fundamental plane
study of Bettoni et al. (2001). The best-fitting relation to 
Bettoni et al. sample is:
\begin{equation}
<\mu>_{e}=3.32(\pm 0.14)\log r_{e} + 16.97(\pm 0.15)
\label{bettoni}
\end{equation}
\noindent
Given that the Bettoni et al. sample is virtually 
at redshift zero, the vertical
offset between the Kormendy relation of the Bettoni et al. sample and
that of the ZP5 sample is a direct indication of the amount of luminosity 
evolution between $z=0$ and $z=0.5$. If the two samples
are fitted with a Kormendy relation of 
fixed slope 3.52 (intermediate to their two
independent fits) then the best-fitting intercepts are
$16.30(\pm 0.07)$ and $16.77(\pm 0.05)$, for the ZP5 and Bettoni et
al. samples respectively. This directly
implies $0.47\pm0.09$  magnitudes of $R-$band luminosity evolution 
between z=0.5 and z=0. This is in excellent agreement 
with the evolutionary corrections derived from the $z_{for}=5$ 
Bruzual \& Charlot model, which predicts a mean value of $0.42\pm0.01$
magnitudes of passive luminosity evolution for the ZP5 sample. 

In the bottom panel of Fig \ref{fig5} the low-redshift radio galaxies
of Bettoni et al. are again plotted, together with the now
evolutionary corrected ZP5 objects. The solid line shows the
best-fitting relation to the combined sample which has the form:
\begin{equation}
<\mu>_{e}=3.35(\pm 0.13)\log r_{e} + 16.93(\pm 0.14)
\end{equation}
\noindent
which is in excellent agreement with the recent determination of the
Kormendy relation slope ($3.33\pm0.09$) for 9000 $z\simeq 0$ early-type 
galaxies drawn from the Sloan Digital Sky Survey (SDSS) by Bernardi et al. (2003). 
\begin{figure*}
\centerline{\epsfig{file=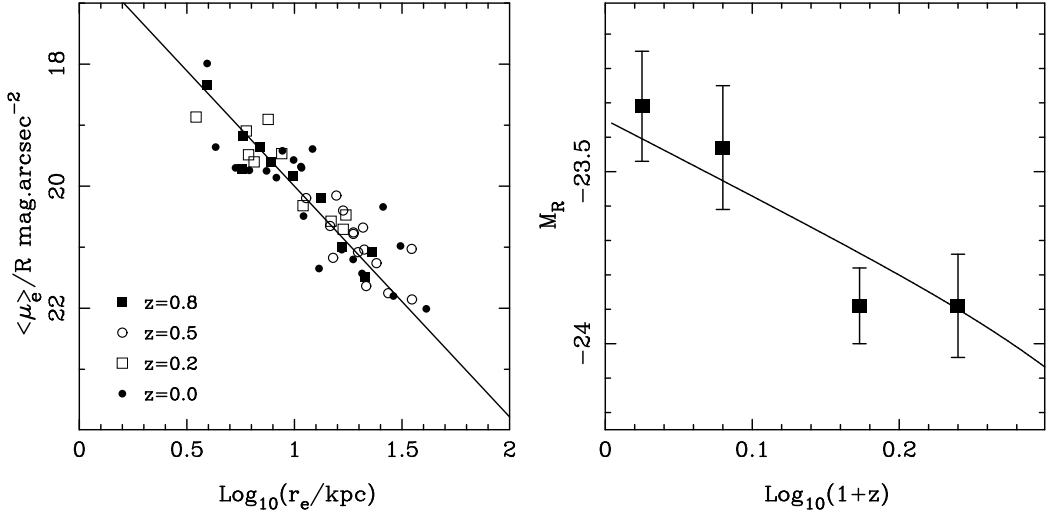,width=14cm,angle=0}}
\caption{Panel A shows the Kormendy relation for the 
four 3C-class radio-galaxy
samples described in the text. The surface brightnesses have been
K-corrected and corrected for passive evolution as described in
Section 5.2. The solid line is 
the best-fitting relation to the combined sample and has a slope 
of $3.78\pm0.30$. Panel B shows the evolution of the mean
absolute $R-$band magnitudes (with standard error bars) of the four 
3C-class samples with redshift, where the absolute magnitudes have
been K-corrected, but
have not been corrected for passive evolution. The solid line shows
the expected evolution of the mean absolute magnitude with redshift
due to the passive evolution of their stellar populations, using the
Bruzual \& Charlot single burst elliptical models with $z_{for}=5$ and
assuming a redshift zero absolute magnitude of $M_{R}=-23.35$.} 

\label{fig6}
\end{figure*}

\section{Evolution of the most powerful radio galaxies}
\label{3c_evol}
In this section we focus on our new results for the 3CRR ZP5
sub-sample, combining them with previous work in the literature, to
re-examine the issue of the evolution of 3C-class radio galaxies from
$z\simeq 0.8$ through to the present day. It has been known for some twenty
years now that powerful radio galaxies follow a tight correlation
between their apparent $K-$band magnitudes and redshift (Lilly \&
Longair 1984). The simplest interpretation of this $K-z$ relation consistent
with the data assumes that the radio galaxies were formed
at high redshift ($z_{for}>3$), either in a monolithic collapse or via
hierarchical merging, and have been evolving passively
thereafter. 

However, this interpretation of the $K-z$ relation was 
challenged by Best, Longair \& R\"{o}ttgering (1998) who argued 
that the apparent passive appearance of the $K-z$
relation was in fact a conspiracy. Based on evidence in the literature
that points to higher redshift 3C radio galaxies ($z\ge 0.5$) inhabiting
significantly richer cluster environments (Hill \& Lilly 1991,
Best 2000) than their low-redshift counterparts 
(eg. Prestage \& Peacock 1988; Lilly \& Prestage 1987), Best et
al. proposed that 3C radio galaxies at $z\simeq 1$ and $z\simeq 0$
followed different evolutionary histories, and are seen as powerful
radio sources at points within those histories at which
they possessed similar stellar masses ($\simeq 10^{11}\Msun$). Under
this interpretation, higher redshift 3C galaxies would form
first, via merger events within their rich cluster environments,
reaching the required stellar mass by $z\simeq 1$, before evolving 
passively to form the radio-dormant brightest cluster galaxies (BCGs) 
seen at $z\simeq 0$. In
contrast, the 3C radio galaxies at low redshift would reach the
required stellar mass much later, via merging within their poorer
cluster environments. The argument in favour of this interpretation of
the $K-z$ relation was further strengthened by Best et al.'s results
suggesting that the scalelengths of $z\simeq 1$ 3C galaxies were
significantly larger than seen in low-redshift 3C galaxies, and much more
compatible with the characteristic sizes of low-redshift BCGs.

The evidence suggesting that 3C radio galaxies at $z\simeq1$ have significantly
larger scalelengths than typical of low-redshift 3C galaxies was
re-examined using a full two-dimensional analysis of the Best et
al. HST imaging data by McLure \& Dunlop (2000). Comparison of the 
results of this analysis with previous results based on HST
imaging of 3C-class radio galaxies at $z\simeq 0.2$ (Dunlop et al. 2003;
McLure et al. 1999) indicated that 3C galaxies at both redshifts had
very similar characteristic scalelengths ($11\pm 2$ kpc; $H_{0}=50,
\Omega_{0}=1$). Here we can extend this comparison by including 
both the new results for the ZP5 3CRR sub-sample, and those for the 
twenty 3C galaxies included in the Bettoni et al. (2001) sample.

\subsection{Scalelengths}

The scalelength results for the $z\simeq0$ sub-sample (Bettoni et
al. 2001), and the $z\simeq0.2$ and $z\simeq0.8$ sub-samples (McLure
\& Dunlop 2000) were all derived by fitting classical $\beta=0.25$ 
de Vaucouleurs $r^{1/4}$ surface-brightness
distributions. Consequently, in order to perform the scalelength
comparison we also adopt the scalelength results for the ZP5 3CRR
sub-sample obtained by restricting the fitting procedure to $\beta=0.25$.

Under this restriction the mean scalelength of the
ZP5 3CRR sub-sample is $16.3\pm 1.6$ kpc. When converted to 
our adopted cosmology the equivalent values for the $z\simeq0$,
$z\simeq0.2$ and $z\simeq0.8$ samples are $14.2\pm 2.2$ kpc, $9.8\pm
1.6$ kpc and $11.4 \pm 2.2$ kpc respectively. In conclusion, these
results show no indication of any trend for scalelength to 
vary with redshift in the interval $0.0<z<0.8$.

\subsection{Evolution of the Kormendy relation}

In panel A of Fig \ref{fig6} we show the Kormendy relation for all four 3C
radio-galaxy samples. The best-fitting relation to the combined sample 
is:
\begin{equation}
<\mu>_{e}=3.78(\pm 0.30)\log r_{e} + 16.22(\pm 0.33)
\end{equation}
\noindent
where all four samples have been corrected for
$(1~+~z)^{4}$ surface-brightness dimming, and have had both K-corrections and
evolutionary corrections applied as described in Section
\ref{lum}.

Panel A of Fig \ref{fig6} demonstrates that 
within the redshift interval $0.0<z< 0.8$ there is no obvious 
evidence for any dynamical evolution, as 
would be indicated by scalelength or surface-brightness evolution, 
other than that expected from purely passive evolution of the stellar
populations. However, although we find no evidence for dynamical
evolution of the class of AGN host galaxies which harbour powerful 
3C-class radio sources in the redshift interval $0.0<z< 0.8$, 
dynamical evolution of the individual host galaxies themselves 
within this redshift range cannot be ruled out.
Panel B of Fig \ref{fig6} shows that the redshift evolution of the mean 
absolute magnitudes of the four sub-samples is also consistent with 
purely passive evolution. However,
it should also be pointed out that these results do not
require that all 3C radio galaxies are formed in a monolithic collapse at high
redshift, simply that the proto-galactic components which eventually
form the host galaxies are formed reasonably co-evally, and that the
vast majority of merger activity be completed at
$z>1$. Furthermore, neither are the results presented in this
section inconsistent with what would be expected under the
$K-z$ interpretation proposed by Best et al. (1998). 

A satisfactory discrimination between the two interpretations of the
$K-z$ relation will rely on a consistent evaluation of the relative
richness of the respective cluster environments at low and high
redshift. Unfortunately, much of the work which has previously been
done on the cluster environments of radio galaxies, particularly at
low redshift, has relied on statistical determinations from
single-filter imaging which are known to be subject to large
uncertainties (eg. Barr et al. 2003). However, upon completion 
of our INT Wide Field Camera
follow-up observations of the ZP5 sample, together with our existing
observations of radio-galaxy samples at $z\simeq0.2$ and $z\simeq0.8$, 
we will be able to consistently investigate the evolution in the cluster
environments of powerful radio galaxies, by using 
three-colour imaging to successfully separate true cluster members 
from foreground and background contaminants (Jarvis et al.; 
McLure et al., in prep).

\section{Comparison with brightest cluster galaxies}
\label{bcg_comp}
\begin{figure}
\centerline{\epsfig{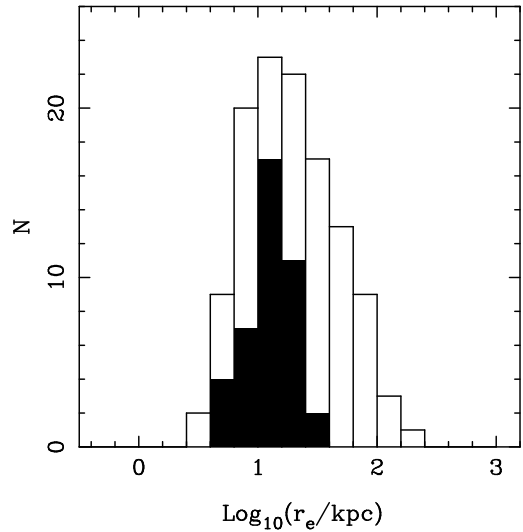}}
\caption{Scalelength histogram for the 119 Abell
brightest cluster galaxies from Graham et al. (1996). Overlaid in
black is the scalelength distribution of the full ZP5 sample.}
\label{fig7}
\end{figure}

The results regarding the scalelengths and integrated luminosities
of the ZP5 sample presented in Section 5 suggest that 
the host galaxies of the ZP5 sample are directly
comparable with brightest cluster galaxies (BCGs). This is in
good agreement with the results of Dunlop et al. (2003) and McLure
\& Dunlop (2001b) who concluded that the hosts of powerful radio-loud
AGN at $z\simeq0.2$ were directly comparable to the central BCGs of
Abell class 0 clusters. In this section we explore this issue further via 
comparison with the findings of the HST imaging study of $z\simeq0.5$
BCGs by Nelson et al. (2002), and the ground-based $R-$band surface 
photometry of 119 local Abell cluster BCGs by Graham et al. (1996). 

Nelson et al. (2002) investigated the characteristic sizes of a
heterogeneous sample of 16 BCGs, drawn from both X-ray and optically
selected cluster samples, using two-dimensional modelling of HST
NICMOS and WFPC2 
imaging with the publically available GIM2D package (Simard 1998). The 
mean redshift of the Nelson et al. sample is $<z>=0.54 \pm 0.03$,
making it ideal for comparison with the results for the ZP5
sample. After conversion to our chosen cosmology, and correction for
the effect of colour gradients (Nelson et al. 2002, Pahre 1999), we
calculate the mean effective scalelength of the Nelson et al. BCGs to
be $<r_{e}>=14.3\pm2.0$ kpc, in excellent agreement with the
scalelength distribution of the ZP5 sample;
$<r_{e}>=13.9\pm0.9$ kpc. Given that the results for both the ZP5 and
Nelson et al. samples are derived from similar two-dimensional modelling of
HST imaging, of samples with identical redshift distributions, this can
be taken as good evidence that the hosts of the ZP5 sample are 
indistinguishable from those of BCGs at $z\simeq0.5$.

However, if the ZP5 scalelength results are compared to 
those determined for nearby Abell BCGs, the conclusions are 
more ambiguous. In Fig \ref{fig7} we
show a histogram of the effective
scalelengths determined from one-dimensional $r^{1/4}$ fits to 
the $R-$band surface-brightness profiles of 119 $z\simeq0$ Abell
cluster BCGs by Graham et al. (1996), together with the equivalent
histogram for the ZP5 sample. The mean and median scalelengths
determined by Graham et al. for the Abell BCGs are $28.7\pm 2.7$ kpc and
18.5 kpc respectively (after conversion to our adopted
cosmology). The equivalent figures for the ZP5
sample are $13.9\pm 0.9$ kpc and 12.8 kpc respectively. An 
application of the Kolmogorov-Smirnov (KS) test finds the ZP5 and Abell
cluster scalelength distributions to be drawn from 
different parent populations at high significance ($p=7\times 10^{-4}$)

\begin{figure}
\centerline{\epsfig{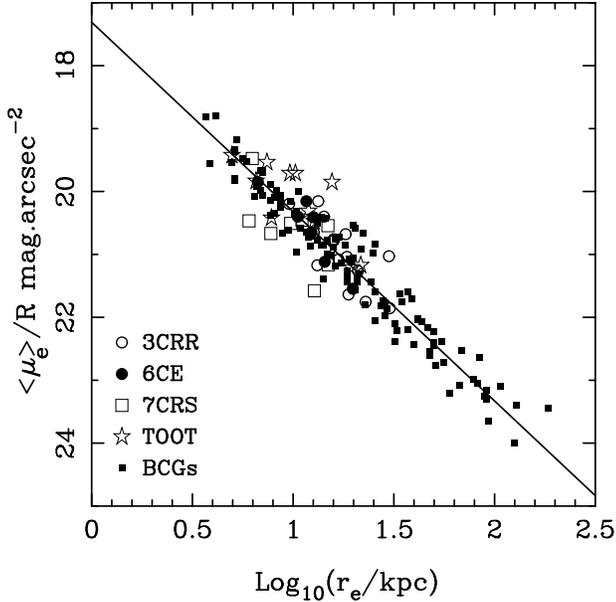}}
\caption{The combined Kormendy relation for the 119 Abell brightest
cluster galaxies of Graham et al. (1996) and the full ZP5 sample. The
solid line is the best-fitting relation to the combined sample and has
a slope of $3.01\pm0.13$.}
\label{fig8}
\end{figure}

In Fig \ref{fig8} the Kormendy relation for the ZP5 sample and 
the Abell BCGs is shown. It is clear from Fig \ref{fig8} that the 
host galaxies of the ZP5 sample are fully consistent 
with the Kormendy relation for $z\simeq0$
Abell BCGs, although the ZP5 sample, and $z\simeq 0.5$ BCGs in
general (Nelson et al. 2002), appear to be devoid of galaxies with 
scalelengths $\gtsim\,~40\,$ kpc. However, the radio galaxies comprising
the ZP5 sample are expected to inhabit a wide range of environments
(Hill \& Lilly 1991), with
previous results suggesting that the mean environmental richness
should be comparable to low-redshift Abell class 0 clusters. Indeed,
if the scalelength comparison with the Graham et al. Abell BCGs is 
restricted to only those objects in Abell class 0 clusters (67 of 119 objects) 
then the difference between the two distributions is only significant 
at the $2\sigma$ level ($p=0.034$), in agreement with 
Hill \& Lilly (1991) and McLure \& Dunlop (2001b).

In conclusion, the luminosities and scalelengths of the ZP5 sample
show that the radio galaxies are consistent with the typical properties of
BCGs at $z\simeq 0.5$, which are in turn comparable with the central
galaxies of local Abell class 0 clusters. We note here that, as
previously highlighted by Yates, Miller \& Peacock (1989), 3C295 is
clearly an exceptional object in terms of it's cluster environment,
which is known to be directly comparable to the richest 
clusters at low redshift (Hill \& Lilly 1991). Indeed, 3C295 is a
factor of two more optically luminous than any other member of the 3CRR
sub-sample and, as will be seen in Section 9, also inhabits an extreme
region of the $L_{151MHz}-M_{bh}$ plane.

\section{The black-hole mass distribution}
\label{black}

\begin{figure}
\centerline{\epsfig{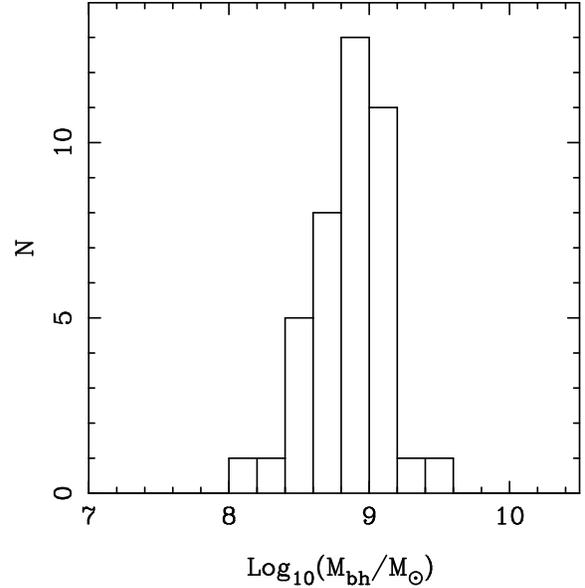}}
\caption{The black-hole mass distribution for the full ZP5 sample. The 
black-hole masses have been estimated from the fitted bulge luminosities via 
the $M_{bh}-M_{bulge}$ relation 
derived by McLure \& Dunlop (2002).} 
\label{fig9}
\end{figure}

As discussed in Section \ref{results}, the morphological analysis of 
the HST imaging data clearly indicates that the ZP5 radio galaxies
are close to being classical ellipticals. Furthermore, it is
possible to directly link the hosts of the ZP5 sample with $z\simeq 0$
ellipticals via the application of passive evolutionary corrections. 
Consequently, under the assumption that, except for passive evolution,
 the $M_{bh}-M_{bulge}$ relation does not evolve between $z=0$ and
 $z=0.5$ we can proceed to estimate the 
central black-hole masses of the ZP5 sample using the correlation 
between black-hole mass and bulge luminosity observed at low
 redshift. 

For this calculation we adopt the $M_{bh}-M_{bulge}$
relation as derived by McLure \& Dunlop (2002) for a combined sample
of 72 AGN host galaxies and 18 low-redshift ($z<0.05$) inactive
ellipticals with black-hole mass measurements derived from either 
stellar or gas dynamics:
\begin{equation}
\log M_{bh}/\Msun = -0.50(\pm0.02)M_{R} -2.74 (\pm 0.48 )
\end{equation}
\noindent
where $M_{R}$ is the absolute R-Cousins bulge luminosity (corrected to
our adopted cosmology). It was 
demonstrated by McLure \& Dunlop (2002) that the scatter associated
with this relation for elliptical galaxies is entirely
consistent with that of the $M_{bh}-\sigma$ relation, at only 0.3 dex
(see also Erwin et al. 2003). Recently, Marconi \& Hunt (2003) have 
performed detailed surface-brightness decompositions of 2-Micron All Sky
Survey (2MASS) imaging of 37 low-redshift (distance $\ltsim$ 150 Mpc)
galaxies with published
black-hole mass determinations, investigating the level of scatter in the
$M_{bh}-M_{bulge}$ relation in the near-infrared. Marconi \& Hunt
confirm the McLure \& Dunlop (2002) results, extending them to
demonstrate that the scatter associated with the $M_{bh}-M_{bulge}$
relation is $\simeq 0.3$ dex, regardless of morphological type,
provided that accurate disc/bulge decompositions of the galaxy
surface-brightness distributions are adopted. The distribution of 
black-hole masses for the full ZP5 sample as estimated from their bulge
luminosities are shown in Fig \ref{fig9}.

The geometric mean black-hole mass for the ZP5 sample 
is $10^{8.87\pm0.04}\Msun$. This is in good agreement with the 
results of Bettoni et al. (2003) who determined a geometric mean 
black-hole mass of $10^{8.81\pm 0.06}\Msun$ and 
$10^{8.91\pm0.06}\Msun$ for 45 low-redshift radio galaxies 
($z\,\ltsim\, 0.1$) using the $M_{bh}-\sigma$ and 
$M_{bh}-M_{bulge}$ relations respectively. When combined with the
results of previous studies at lower redshift (eg. Laor 2000; McLure \& Dunlop
2002; McLure \& Jarvis 2002; Dunlop et al. 2003; Bettoni et al. 2003;
Barth et al. 2003) it is clear that truly powerful radio-loud AGN,
those capable of producing kpc-scale jets, are powered by black holes
with masses greater than $10^{8}\Msun$. In addition, given that
radio-loud AGN are drawn from the high-mass end of the black-hole mass
function, the results of these studies suggest the existence of a
maximum black-hole mass of $\simeq 3\times 10^{9}\Msun$
(c.f. 3C295). This black-hole mass limit is in agreement with the
maximum mass measured via gas or stellar dynamics in the local
Universe (Ford et al. 1994; Tadhunter et al. 2003) and the limit
implied from broad emission-line based black-hole mass estimates 
for $>12000$ SDSS quasars at $z\leq2$ (McLure \& Dunlop 2004).

\subsection{On the correlation between $\beta$ and black-hole mass}
\label{graham}

We note here that Graham et al. (2001) discovered a correlation between
central black-hole mass and the S\'{e}rsic morphological parameter
$\beta$ for a sample of 20 local galaxies with black-hole masses 
measured via gas or stellar dynamics. This correlation displays 
a comparable level of scatter ($\sim 0.3$ dex) to that of the $M_{bh}-\sigma$
relation (eg. Tremaine et al. 2002). Due to the fact that black-hole
mass in early-type galaxies also correlates with bulge luminosity 
with comparably low scatter (McLure \& Dunlop 2002; Marconi \& Hunt
2003), it is expected that a correlation will exist between the $\beta$
parameter and host-galaxy luminosity.

No such correlation is observed in the full ZP5 sample ($r_{s}=0.20,
p=0.21$). It is possible that the properties of two of the ZP5 sample, 
TOOT 0013+3459 and 3C295 (see Table 3), may have resulted in the
determination of anomalously large $\beta$ parameters for 
their luminosities. However, following the exclusion of these two
objects the correlation between $\beta$ and host luminosity is still
only marginally significant ($r_{s}=0.40, p=0.012$, $\simeq 2.4\sigma$).
The principal reason for this discrepancy may simply be that we 
have a much smaller dynamic range in $\beta$ compared to the 
Graham et al. (2001) study. However, we also
note that the Graham et al. correlation between $\beta$ and black-hole
mass would imply a mean black-hole mass for the ZP5 sample of
$10^{8.4}\Msun$, for our determination of $<\beta>=0.23\pm 0.01$. This is
significantly different to the figure of $10^{8.87\pm 0.04}\Msun$
determined from the $M_{bh}-M_{bulge}$ relation which, using the
Graham et al. $\beta-M_{bh}$ correlation, would imply $<\beta>=0.16$.

\section{The black hole mass - radio luminosity connection}
\label{correlations}
\begin{figure*}
\centerline{\epsfig{file=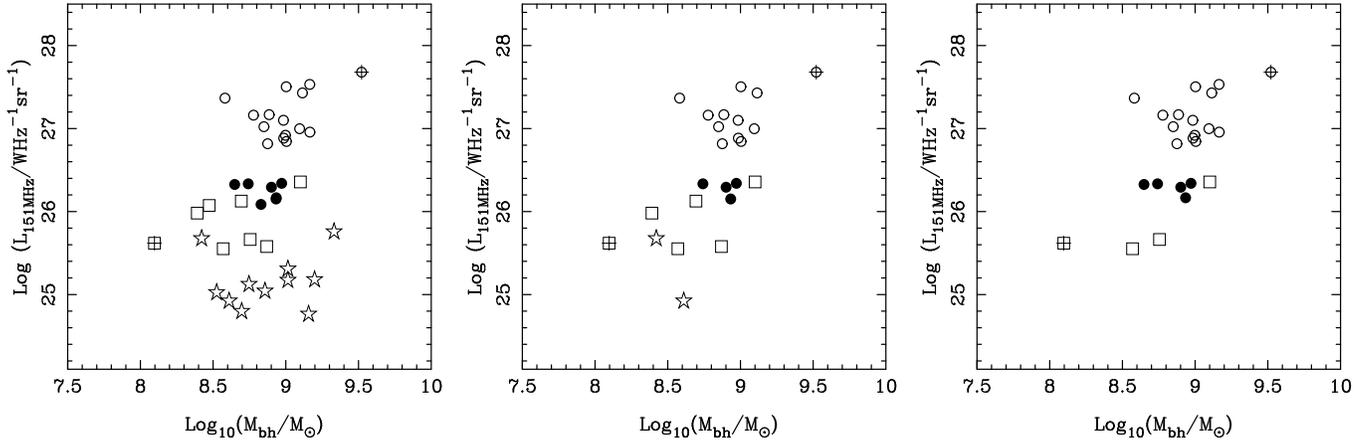,width=18cm,angle=0}}
\caption{Panel A shows 151-MHz radio luminosity plotted against
black-hole mass for the full ZP5 sample. Panel B shows the
$L_{151}-M_{bh}$ plane for the ZP5 objects with HEG spectra. Panel C
shows the $L_{151}-M_{bh}$ plane for the ZP5 objects with classical
double (CD) radio structures. The black-hole masses have been
estimated from the bulge luminosity fits using the 
$M_{bh}-M_{bulge}$ correlation of McLure \& Dunlop (2002). The two 
objects highlighted with crosses are 7CRS1731+6638 and 3C295 which 
possess the two extreme values of black-hole mass and have a tendency
to lead the eye. However, these two objects are included in the
correlation analysis unless stated otherwise.}
\label{fig10}
\end{figure*}

The question of whether or not black-hole mass is correlated with
radio emission in AGN has been the focus of a great deal of attention
in the literature recently (eg. Dunlop et al. 2003; Lacy et al. 2001; 
Ho 2002; Woo \& Urry 2002). Using broad emission-line based virial
black-hole mass estimates Lacy et al. (2001) investigated the
radio luminosity - black-hole mass relationship for a 
combined sample composed of
objects from the First Bright Quasar (FBQ) and PG quasar surveys. The 
results of this study showed a strong correlation between black-hole
mass and 5-GHz radio luminosity of the form $L_{5GHz} \propto
M_{bh}^{1.4\pm 0.2}$ over a dynamic range of some 6 decades in radio
luminosity and 4 decades in black-hole mass. By combining their
black-hole mass estimates, derived from HST imaging of a sample of 
33 AGN at $z\simeq0.2$, with those of the Lacy et al. study, Dunlop \&
McLure (2003) demonstrated that the 5-GHz radio luminosity of both AGN and
quiescent nearby galaxies appears to be confined between two
radio-power limits. The lower radio-power limit defines the minimum
radio luminosity that can be emitted by a given black-hole mass, and
is well described by
a relation of the form  $L_{5GHz} \propto M_{bh}^{2.5}$, in good
agreement with the correlation originally observed for nearby
quiescent galaxies by Franceschini, Vercellone \& Fabian (1998). The
upper radio-power limit proposed by Dunlop \& McLure (2003) appears to be well
described by a relation of the same functional form, offset from the
lower limit by some 5 decades in radio luminosity. In contrast, the
recent studies of Ho (2002) and Woo \& Urry (2002)
find no convincing evidence for a correlation between black-hole mass
and radio luminosity, in samples comprising a range in nuclear activity
from local quiescent galaxies up to and including powerful quasars. 

In this section we investigate whether there is a correlation between
black-hole mass and radio luminosity within the ZP5 radio-galaxy
sample. In contrast to previous studies, the majority of which have
been based on high frequency (5-GHz) radio luminosity, the ZP5
radio-galaxy sample allows us to test for a correlation between 
black-hole mass and extended low-frequency (151-MHz) radio
luminosity. This distinction is potentially important given that
$L_{151MHz}$ is less affected by beaming than $L_{5GHz}$ (eg. Jarvis
\& McLure 2002) and has a close relationship to the time-averaged
kinetic energy of the jets (e.g. Rawlings \& Saunders 1991).

The results presented thus far are consistent with a
picture in which extended low-frequency radio luminosity scales roughly
with host-galaxy luminosity/mass. In combination with the latest 
determination of the $K-z$ relation (Willott et al. 2003) it appears 
that the most powerful 3C-class radio galaxies
reside in galaxies with $R-$band luminosities of $\simeq 4L^{\star}$,
while the lower-luminosity 6C and 7C-class radio galaxies typically
inhabit hosts with luminosities of $\simeq 3L^{\star}$ and 
$\simeq 2L^{\star}$ respectively.

At present it is unclear how the TOOT sub-sample fits within this
picture. It can be seen from the results presented in
Table \ref{tab4} that the TOOT galaxies do not appear to follow the
rough scaling between extended radio luminosity and host-galaxy
luminosity apparent in the 3CRR, 6CE and 7CRS sub-samples. Indeed, the
mean luminosity of the TOOT sub-sample ($3.28\pm0.54 L^{\star}$) is 
greater than that of both the 7CRS and 6CE sub-samples, and is
consistent with the low-redshift results of Owen \& Laing (1989) that fat
double/jet/FRI sources reside in hosts which are on average $\simeq
0.5$ magnitudes brighter than those of classical double/FRII sources
of comparable radio luminosity. However, as was mentioned in 
Section 2, the TOOT sub-sample was drawn from a
preliminary version of the survey and it is unclear at the time of
writing to what extent the current TOOT sub-sample is biased by the
exclusion of the optically faintest sources.

With this in mind, panel A of Fig \ref{fig10} shows 151-MHz radio
luminosity versus estimated black-hole mass for the full ZP5 sample,
where the black-hole mass estimates have been derived via the McLure
\& Dunlop (2002) $M_{bh}-M_{bulge}$ relation as described in Section
8. Taken as a whole, there is only a weak ($r_{s}=0.35, p=0.027$, $2.2\sigma$)
correlation between black-hole mass and radio luminosity 
within the ZP5 sample. However, as suggested previously, it can be 
seen from panel A of Fig \ref{fig10} that the apparent weakness of the
$L_{151}-M_{bh}$ correlation displayed by the full ZP5 sample is due,
at least in part, to the inclusion of the eleven TOOT objects.

To investigate the possible influence of radio structure and nuclear
spectral type we have also plotted in Fig \ref{fig10} the $L_{151}-M_{bh}$ 
relation for two sub-sets of the ZP5 sample. In panel B of Fig
\ref{fig10} we show the $L_{151}-M_{bh}$ relation for those objects
which display high-excitation nuclear spectra (HEG) only. As is suggested
by the figure, this sub-sample displays a fairly strong correlation which is
significant at the $99.9\%$ level ($r_{s}=0.62, p=0.001$,
$3.0\sigma$), although this drops in significance to only $2.4\sigma$
following the exclusion of 3C295 and 7CRS 1731+6638 (objects
highlighted with crosses in Fig \ref{fig10}). In panel C of 
Fig \ref{fig10} we show the $L_{151}-M_{bh}$ relation
for the ZP5 objects with classical double (CD) radio structures 
only, effectively excluding the TOOT sub-sample. This correlation is
significant at the $2.7\sigma$-level ($r_{s}=0.56, p=0.005$), a result
which is confirmed by the application of the generalized
Kendall's tau test ($\tau=0.83$, $2.8\sigma$), although again we note that
the significance drops to only $\simeq 2\sigma$ ($r_{s}=0.43,p=0.049$)
following the exclusion of 3C295 and 7CRS 1731+6638.

It is clear from these results that a $\simeq 3\sigma$ correlation
does exist between black-hole mass and low-frequency 151-MHz radio    
luminosity within the 3CRR, 6CE and 7CRS samples, although the scatter
in radio luminosity at a given black-hole mass is large ($\simeq
1.5$ dex). This result is in good agreement with the 
Lacy et al. (2001) study of the $L_{5GHz}-M_{bh}$ relation in 
broad-line quasars, a study which covered a much larger dynamic
range in black-hole mass. Moreover, this result is also consistent
with the hypothesis proposed by Dunlop \&
McLure (2003) that the radio luminosity which can be produced by a 
given black-hole mass may be bounded by upper and lower limits 
separated by several orders of magnitude. Indeed, given that 
theoretical models indicate that the radio luminosity of AGN, in 
addition to black-hole mass, should also be related to both 
accretion rate and black-hole spin (eg. Meier 2003), it is perhaps 
unsurprising that the $L_{151}-M_{bh}$ correlation is not stronger.

Furthermore, at present
it is unclear whether the objects which do not follow the $L_{151}-M_{bh}$
correlation, predominantly the lower-luminosity TOOT sub-sample, do so
because of their LEG nature, their fat double/jet radio structures, or 
alternatively, as a result of a selection effect which biases 
the TOOT sub-sample towards the inclusion of objects with larger 
host luminosities/black-hole masses. A more detailed analysis of the 
connection between the radio properties and black-hole masses of 
the ZP5 sample will be pursued in a future paper (Mitchell et al., in prep).

\section{Conclusions}

The results of a detailed analysis of deep, high-resolution HST images
of a sample of 41 radio galaxies which spans three decades in radio
luminosity at $z\simeq 0.5$ have been presented. The host-galaxy
properties derived from two-dimensional image modelling have been
discussed and compared to existing results in the literature. The
principle conclusions of this study can be summarized as follows:
\begin{enumerate}

\item{The morphologies of the host galaxies of all four radio-galaxy 
sub-samples are found to be largely consistent with a 
canonical elliptical-galaxy
$r^{1/4}$ surface-brightness distribution. No trend is detected for
host-galaxy morphology to vary significantly with radio luminosity.}

\item{The luminosities of the host galaxies are found to be consistent
with those of galaxies drawn from the bright end of the local 
cluster-galaxy luminosity function, in the range 
$0.7L^{\star}<L<10L^{\star}$, with a mean value of $3.2\pm0.3 L^{\star}$.}

\item{The radio-galaxy hosts are found to lie on the same Kormendy
($\mu_{e}-r_{e}$) relation as powerful $z\simeq 0$ radio galaxies, after
allowing for the effects of passive evolution of their stellar
populations under the assumption of a high formation redshift 
($z_{for}\,\gtsim$\,3).}

\item{By combining the new results on $z\simeq 0.5$ 3CRR galaxies with 
previous results in the literature, it is found that there is no
evidence for dynamical evolution among the hosts of 3C-class radio
galaxies in terms of scalelength, luminosity and Kormendy relation in 
the redshift interval $0.0<z<0.8$. However, although no evidence is
found for dynamical evolution among the hosts of the most powerful 
radio sources as a class, dynamical evolution of the individual host 
galaxies themselves cannot be ruled out.}

\item{The scalelengths of the ZP5 host galaxies are found to be
consistent with those of brightest cluster galaxies (BCGs) at the same cosmic
epoch ($z\simeq 0.5$). When compared to local Abell cluster BCGs the
ZP5 host galaxies are found to be a factor of $\simeq 1.5$ smaller on
average. However, when the comparison is restricted to local clusters
of Abell class 0 only, the ZP5 and BCG scalelength distributions are
found to be statistically indistinguishable.}

\item{Converting the derived host-galaxy luminosities into black-hole
mass estimates using the local $M_{bh}-M_{bulge}$ relation predicts that the 
host galaxies harbour central black-holes in the 
mass range $10^{8.1}\Msun<M_{bh}<10^{9.5}\Msun$, with 
a geometric mean of $10^{8.87\pm0.04}\Msun$}.

\item{Significant correlations ($\simeq 3\sigma$) are found 
between black-hole mass and extended low-frequency radio luminosity
for sub-samples of the ZP5 objects which have either classical double
(CD) radio structures or high-excitation nuclear spectra (HEG). However,
at present we are unable to determine whether the objects which 
do not follow the same correlation, predominantly the lower-luminosity
TOOT sub-sample, do so because of their LEG spectra, fat double/jet 
radio structures, or alternatively, as a result of
selection effects biasing the TOOT sub-sample towards the inclusion of
objects with higher host luminosities/black-hole masses.}

\end{enumerate}

\section*{ACKNOWLEDGMENTS} 
RJM acknowledges the award of a PPARC
postdoctoral fellowship and a PPARC PDRA. CJW thanks the National
Research Council of Canada for support. MJJ acknowledges the support
of the European Community Research and 
Training Network `The Physics of the Intergalactic Medium' and a PPARC 
PDRA. This material is based in part 
upon work supported by the Texas Advanced Research Program 
under Grant No. 009658-0710-1999 and NASA/STScI grant HST-GO-09045.04-A.
EM acknowledges the award of a PPARC
studentship. JSD acknowledges the enhanced research time provided by the
award of a PPARC Senior Fellowship. MW acknowledges the Swedish
Research Council for travel funding to La Palma. Some of the data 
presented in this paper
were obtained from the Multimission Archive at the Space Telescope 
Science Institute (MAST). STScI is operated by the Association of
Universities for Research in Astronomy, Inc., under NASA contract
NAS5-26555. Support for MAST for non-HST data is provided by the 
NASA Office of Space Science via grant NAG5-7584 and by other 
grants and contracts. This research has made use of the 
NASA/IPAC Extragalactic Database (NED)
which is operated by the Jet Propulsion Laboratory, California
Institute of Technology, under contract with the National Aeronautics
and Space Administration.

{}
\end{document}